# Direct evidence for charge compensation induced large magnetoresistance in thin WTe$_2$

Yaojia Wang[∥,†], Lizheng Wang[∥,†], Xiaowei Liu[†], Heng Wu[†], Pengfei Wang[†], Dayu Yan[‡], Bin Cheng[†], Youguo Shi[‡], Kenji Watanabe[§], Takashi Taniguchi[§], Shi-Jun Liang[*,†], Feng Miao[*,†]

[†]National Laboratory of Solid State Microstructures, School of Physics, Collaborative Innovation Center of Advanced Microstructures, Nanjing University, Nanjing 210093, China.

[‡]Institute of Physics, Chinese Academy of Sciences, Beijing 100190, China

[§]National Institute for Materials Science, 1-1 Namiki Tsukuba, Ibaraki 305-0044, Japan

**Abstract:** Since the discovery of extremely large non-saturating magnetoresistance (MR) in WTe$_2$, much effort has been devoted to understanding the underlying mechanism, which is still under debate. Here, we explicitly identify the dominant physical origin of the large non-saturating MR through *in-situ* tuning of the magneto-transport properties in thin WTe$_2$ film. With an electrostatic doping approach, we observed a non-monotonic gate dependence of the MR. The MR reaches a maximum (10600%) in thin WTe$_2$ film at certain gate voltage where electron and hole concentrations are balanced, indicating that the charge compensation is the dominant mechanism of the observed large MR. Besides, we show that the temperature dependent magnetoresistance exhibits similar tendency with the carrier mobility when the charge compensation is retained, revealing that distinct scattering mechanisms may be at play for the temperature dependence of magneto-transport properties. Our work would be helpful for understanding mechanism of the large MR in other nonmagnetic materials and offers an avenue for achieving large MR in the non-magnetic materials with electron-hole pockets.

Keywords: WTe$_2$, large non-saturating magnetoresistance, charge compensation, in-situ tuning

As a layered non-magnetic material, WTe$_2$ exhibits an extremely large non-saturating magnetoresistance (MR) [1]. This discovery has stimulated the observation of topological Weyl state [2-4], superconductivity[5-10], ferroelectricity [11] and quantum spin Hall state in WTe$_2$ [12]. Uncovering and identifying the physical origin of this unusually large MR is of crucial significance for devising novel magnetic sensor and nanostructure memory devices. So far, distinct physical mechanisms, such as electron-hole compensation [13-15] and spin texture induced suppression of backscattering[16, 17], have been proposed to explain the physical origin. The charge compensation as the most studied mechanism has been supported by the observed nearly identical electron and hole pockets at Fermi surface by angle-resolved photoemission spectroscopy (ARPES)[15, 18-20]. However, these ARPES experiments only offer an indirect evidence. The most desirable way to validate this mechanism is to study the magneto-transport property by *in-situ* tuning carrier density across the compensation point. By using some approaches such as high-pressure [21], chemical [22, 23] or electrostatic doping [24] to tune the carrier density, the observed reduction in the MR seems to be consistent with the charge compensation mechanism. Nevertheless, it is challenging to access the charge compensation point in these experiments due to sample degradation or narrow tuning range of the carrier density. Moreover, the variation of carrier mobility during tuning of the carrier density would affect the MR, as pointed out in many experiments in which the low mobility leads to small MR in thin films (*i.e.* less than 1000% in samples with thickness below 10 nm) and a reduction in the MR of bulk samples[25-38]. The entanglement of different factors in the study of MR makes it challenging to identify the intrinsic mechanism. Therefore, the physical origin of the large MR is still under debate and more effort is required to uncover it explicitly.

In this work, we performed a systematic study on *in-situ* tuning of the MR in WTe$_2$ thin film devices (about 10 nm) via an electrostatic doping approach. With hexagonal boron nitride (h-BN) as the protective and dielectric layer, the MR is tuned to exhibit dramatic change with non-monotonic gate dependence with peak value appearing at a certain voltage. Based on theoretical analysis, we demonstrate that the MR reaches the maximum value at the charge compensation point and decreases rapidly with carrier concentration deviating from the compensation point. In addition, the achieved MR (10600%) at charge compensation is one to two orders of magnitude larger than previous reported values (100% ~ 1000% at 14 T) in thin films with similar thickness (~10 nm)[24, 28-34, 36]. These results unambiguously identify the charge compensation effect as the dominant mechanism for the observed large MR in WTe$_2$ in an explicit manner. Furthermore, we show that the charge compensation can be retained within a wide temperature regime where the temperature dependence of the MR shows the similar tendency with carrier mobility, which enables us to reveal the charge scattering mechanisms for the temperature dependence of MR.

The WTe$_2$ devices covered with h-BN were fabricated in an inert atmosphere glove box to avoid quality degradation[26, 33, 35, 39, 40]. Both WTe$_2$ thin films (about 10 nm) and h-BN films (about 20 nm) were first exfoliated onto SiO$_2$/Si substrate via mechanical exfoliation. We then picked up the h-BN and WTe$_2$ films successively and transferred them onto the pre-prepared Hall bar electrodes (Ti 5 nm/Au 30 nm) by using polypropylene carbon (PPC) films [41]. Few-layer graphene (FLG) flakes exfoliated on PDMS were finally transferred onto the h-BN films to act as top gate electrodes. Figure 1a shows the schematic cross-sectional structure and the optical image of a fabricated device. All electrical measurements in this work were performed using the conventional four-probe technique.

We first characterized the magneto-transport properties of a thin film device by measuring the temperature dependence of the longitudinal resistivity ($\rho(T)$) in different perpendicular magnetic fields, with results shown in Figure 1b. The sample shows metallic behavior at zero field. The ratio of residual resistivity ratio RRR ≡ $\rho(300$ K$)/\rho(2$ K$)$ is about 64, indicating the high quality of the h-BN protected thin film devices. At low temperature region (1.6 K < $T$ < 15 K), the temperature dependence of zero field resistivity can be well described by $\rho_{xx}(T, 0) = \rho_0 + aT^2$ (as shown in the inset) manifesting electron-electron scattering at low temperatures in WTe$_2$ [42], where $\rho_0$ is the residue resistivity and $a$ is a free parameter. With applying a magnetic field, $\rho_{xx}(T, B)$ curves present a transition from metallic-like state at high temperatures to insulator-like state at low temperatures. This is similar to the turn-on behavior observed in bulk WTe$_2$ [42-44], suggesting the onset of large MR at low temperatures in our high-quality thin film devices. Such turn-on behavior is not the signal of field induced metal-insulator phase transition in WTe$_2$, but a manifestation of the MR behavior determined by the two-band theory. Figure 1c shows the MR measured at 1.6 K, where $\rho_{xx}(B)$ exhibits non-saturating increase with magnetic field and Shubnikov-de Haas (SDH) oscillations at high field. The MR(B) curve (MR(B) = ($\rho_{xx}(B)$ - $\rho_0$)/$\rho_0$ × 100%) can be fitted by a power law of MR ~ $B^n$ with $n$ = 1.81 (Inset of Figure 1c), close to the quadratic relation in bulk WTe$_2$ [42, 45, 46].

By applying the electrostatic doping through h-BN film, we can continuously tune the magneto-transport properties of the thin WTe$_2$ film, which offers a direct way to reveal the physical origin of the large non-saturating MR. Figure 2(a) shows the gate voltage dependence of longitudinal resistivity measured with and without magnetic field at 1.6 K. Compared to the resistivity at zero field, the longitudinal resistivity at a finite magnetic field ($B$ = 12 T) varies non-monotonously with gate voltage ($V_{tg}$) and reaches the maximum at certain $V_{tg}$. To gain further insight into the gate voltage dependence of the MR, we present the MR and the Hall resistivity at different gate voltages, as shown in Figure 2(b) and Figure 2(c) respectively. The MR curve exhibits a non-monotonic change with increasing gate voltage from -10 V to 6 V. The $\rho_{xy}(B)$ curves show a nonlinear feature, indicating the existence of two types of charge carriers in the device. At high field, the Hall resistivity changes dramatically with the gate voltage, and presents a sign change near the voltage of -6 V, which indicates the significant change of charge carriers.

The relation between the MR and charge carriers can be analyzed quantitatively by using the two-band theory [47], which can be described by:

$$\rho_{xx}(B) = \frac{(n_e\mu_e+n_h\mu_h)+(n_e\mu_h+n_h\mu_e)\mu_e\mu_h B^2}{e[(n_e\mu_e+n_h\mu_h)^2+(n_h-n_e)^2\mu_e^2\mu_h^2 B^2]}, \quad (1)$$

$$\rho_{xy}(B) = \frac{(n_h\mu_h^2-n_e\mu_e^2)B+(n_h-n_e)\mu_e^2\mu_h^2 B^3}{e[(n_e\mu_e+n_h\mu_h)^2+(n_h-n_e)^2\mu_e^2\mu_h^2 B^2]}, \quad (2)$$

$$\text{MR}(B) = \frac{(n_e\mu_e+n_h\mu_h)^2+\mu_e\mu_h(n_e\mu_e+n_h\mu_h)(n_h\mu_e+n_e\mu_h)B^2}{(n_e\mu_e+n_h\mu_h)^2+(n_h-n_e)^2\mu_e^2\mu_h^2 B^2} - 1. \quad (3)$$

Where $n_e(n_h)$ is the electron (hole) density, $\mu_e(\mu_h)$ is the electron (hole) mobility, $e$ is the elementary charge. At each voltage, the MR($B$) and $\rho_{xy}(B)$ curves are fitted simultaneously by Eqs. (3) and (2), with fitting curves (dashed lines) plotted in Figure 2b and Figure 2c. The excellent agreement between the fitted curves and experimental data indicates that the two-band theory can be used to account for the varying behavior of MR. Thus, we can further extract the carriers' mobilities and densities at different gate voltages, with the results shown in Figure 2d and Figure 2e respectively. With increasing gate voltage, the electron mobility ($\mu_e$) decreases but hole mobility ($\mu_h$) increases (Figure 2d). The oppositely varying trends of electron and hole mobilities may originate from the surface scattering induced nonlocal effect [30]. Applying a negative (positive) gate voltage would increase (reduce) the surface scattering of holes and lead to the reduction (increase) of the hole mobility, which has opposite effects on electrons. Changing the gate voltage from -10 V to 6 V also gives rise to an opposite variation in the electron ($n_e$) and hole densities ($n_h$). This is reasonable as applying a positive (negative) gate voltage will accumulate electrons (holes) in the thin film. It is noticeable that, with increasing gate voltage, a crossover occurs near -6 V, indicating the dominant carrier changes from hole to electron after passing this crossover point. This carrier type changing behavior is well consistent with the sign change of the Hall resistivity near $V_{tg}$ = -6 V (Figure 2c), which can be interpreted by the two-band theory. According to Eq. (2), the sign of Hall resistivity is mainly determined by the terms $(n_h\mu_h^2-n_e\mu_e^2)B$ and $(n_h-n_e)\mu_e^2\mu_h^2B^3$ at relatively low and high field, respectively. Therefore, the sign of Hall resistivity at high enough field will change when the dominate carrier type varies.

*In-situ* tuning of carrier concentration across the compensation point enables us to establish a direct and explicit relation between the large MR and the charge compensation. We extracted the MR values at $B = 14$ T and calculated the ratio of carrier densities ($n_e/n_h$) at different gate voltages, with results shown in Figure 2f. When $V_{tg}$ changes from -10 V to 6 V, the ratio $n_e/n_h$ increases monotonically from 0.87 to 1.39. In particular, the MR - $V_{tg}$ curve shows a non-monotonic feature and reaches the maximum at the charge compensation point where $n_e/n_h$=1. The MR value decreases rapidly when carrier concentration slightly deviates from the perfect charge compensation point. As the carrier mobility changes during the gate tuning process (Figure 2d), we also compared the gate dependence of $\mu_e\mu_h$ with the MR - $V_{tg}$ curve (see Supporting Information). With the opposite gate dependence of electron and hole mobilities, the change of $\mu_e\mu_h$ is small and its varying trend is also inconsistent with the variation of MR, indicating that the carrier mobility is not the origin of the unusual change of MR. Moreover, the large MR value (~10600%) in our thin film (about 10 nm in thick) is about one to two orders of magnitude larger than previous reported values in samples with similar film thickness[24, 28-34, 36]. Even, this value is also much larger than that in thicker films with higher mobility[33, 36]. It is the perfect charge compensation that gives rise to the large MR in our thin film, which has never been achieved in thin film devices. In addition, we also observed the non-saturated increase of MR(B) curves near the charge compensation point, but the increasing rate of MR(B) becomes small at high field when $n_e/n_h$ is far away from 1 (as shown in Supporting Information). These results unambiguously demonstrate that the charge compensation is the dominate physical origin of the large non-saturating MR in WTe$_2$.

Another important feature of the MR in WTe$_2$ can be manifested by its strong dependence on temperature. This behavior has been proposed to arise from the temperature-dependent electronic band structure of WTe$_2$ [15, 43, 48]. However, recent works [49, 50] have shown that the temperature has a negligible effect on the band structure of WTe$_2$. The achieved charge compensation in our samples also allows us to explicitly study the effect of temperature on the MR and carrier dynamics. We carried out measurements of the MR and the Hall resistivity at different temperatures with $V_{tg}$ = -6 V. As shown in Figure 3a, increasing temperature leads to the suppression of MR and a transition from nonlinear to quasi-linear curves of the Hall resistivity. We extracted the MR values at different magnetic fields and presented them as a function of temperature, as shown in Figure 3b. Noticeably, the MR(T) curves show similar tendency under different magnetic fields. The MR becomes saturated at low temperatures but dramatically decreases when $T > 10$ K. This is distinct from the field dependent turn-on behavior of the resistivity curves shown in Figure 1b, which further indicates that the MR(T) curves reflect the intrinsic characteristic of the magneto-transport properties in WTe$_2$ instead of the turn-on behavior of $\rho(T)$ curves[42, 43].

We further employed two-band theory to analyze the experimental data shown in Figure 3a (see the fitting results in the Supporting Information). The extracted carrier density and mobility at different temperatures are presented in Figure 3c and Figure 3d, respectively. We found that both electron and hole concentrations increase slightly at low temperatures but always keep a nearly-perfect charge compensation for T ≤ 50 K (inset of Figure 3c). At higher temperatures, an opposite variation behavior of electron and hole concentrations was observed, suggesting a thermal excitation induced change of Fermi surface [48]. Different from the variation of electron and hole densities, the temperature dependence of electron mobility is similar to hole mobility. They are nearly independent of temperature below 10 K and start to decrease above 10 K. This temperature dependent behavior is similar to the tendency of MR($T$) curves in Figure 3b.

By keeping charge-compensation over a wide temperature range, it is helpful for revealing the underlying physics of the temperature dependent magneto-transport. Under the condition of charge compensation, the Eq. (3) can be reduced to MR($T$) = $\mu_e(T)\mu_h(T)B^2$, thus the temperature dependence of the MR is determined only by the change of the mobility under a magnetic field. With T < 50 K, we observed the similar temperature dependent variation of the MR and the mobility, which is consistent with the theory and further implies the same scattering mechanism behind these two different physics behaviors. At 10 K < T < 50 K, the mobility is well fitted by $\mu \propto T^{-n}$ with $n$ = 1.23 and 1.2 for electron and hole respectively, which suggests that the electron-phonon interactions dominate the scattering process within this temperature range [51]. Over the same temperature range, we fitted the temperature dependence of the MR at $B = 14\ T$ by using the equation MR($T$) $\propto (a+bT^{-m})^2$, where $a$, $b$ and $m$ are free parameters. The extracted parameter $m$ is 1.22, close to the obtained values of $n$ for the mobility. This suggests that the dramatic decrease in the MR could be attributed to the electron-phonon scattering mechanism. At low temperatures ($T$ < 10 K), the MR tends to saturate, which is also observed at other gate voltages (Supporting Information). Note that similar saturation behavior has also been reported in bulk WTe$_2$, but the physical origins remain unclear. In our samples, the temperature dependent mobility curves exhibit same plateau at $T < 10$ K, indicating that the charge scattering mechanism dominates the magneto-transport. Together with the evidence that electron-electron (e-e) interactions are at play at low temperatures in the Figure1b, we deduce that the e-e interactions may be the dominant scattering source that leads to the plateau of the MR and the mobility in WTe$_2$ and the impurity scattering at low temperature may play a role. Besides, at $T > 100$ K, both electron-phonon scattering and thermal excitation induced Fermi surface change account for the change in the MR. These results may be helpful for understanding the temperature magneto-transport in other materials with large MR [52-57] like MoTe$_2$, LaSb, WP$_2$, PtBi$_2$, etc.

In conclusion, we studied the physical origin of the large non-saturating MR in WTe$_2$ based on the high-quality thin film device. Through an electrostatic tuning approach, we achieved an *in-situ* tuning of charge carriers across the charge compensation point for the first time. The observed maximum MR with a record-high value (~10600%) at the charge compensation point offers a direct evidence for the charge compensation induced large non-saturating MR in the WTe$_2$. By keeping the charge compensation over a wide temperature range, we clearly identified the close relation between temperature dependent MR and mobility, which further reveals the charge scattering mechanisms responsible for temperature dependence of magneto-transport. Our work opens up an avenue for achieving the large MR in the non-magnetic materials with electron-hole pockets and pave the way towards realization of nanostructure magnetic sensor and memory devices.

**Growth of single crystals**

The high quality single crystals of WTe$_2$ were grown by using high-temperature self-flux method. Both tungsten powders (99.9%) and tellurium pieces (99.999%) with ratio of 1:30 were placed into alumina crucibles in a glovebox full of inert gas, and sealed in quartz tubes under high vacuum. The tubes were heated and maintained at 1373 K for 10 h and then cooled down to 923 K slowly with a rate of 2 K/h. The tungsten flux was separated in a centrifuge at 923 K.

**Electrical measurement**

The thin film device was measured in the Oxford cryostat with magnetic field up to 14 Tesla and a base temperature of about 1.6 K. The resistance was measured by using low-frequency Lock-in amplifier.

## ASSOCIATED CONTENT

**Supporting information**

The Supporting information is available free of charge on the ACS Publications website at http://pubs.acs.org.

Effect of carriers' mobilities on the gate tunable magnetoresistance; the effect of the ratio of carrier density on the MR; two-band theory analysis of the magnetoresistance and the Hall resistivity curves at different temperatures; temperature-dependence of the magnetoresistance at different gate voltages; contact resistance of the film device at 1.6 K.

## AUTHOR INFORMATION

**Corresponding Authors**

* E-mail: sjliang@nju.edu.cn.

* E-mail: miao@nju.edu.cn. Tel: +86-025-83621497. Fax: +86-025-83621497.

**ORCID**

Feng Miao: 0000-0002-1910-9781

Bin Cheng: 0000-0002-2932-4370

Kenji Watanabe: 0000-0003-3701-8119

**Author Contributions**

∥Y.W. and L.W. contribute equally to this work.

**Notes**

The authors declare no competing financial interest.

## Acknowledgements

This work was supported in part by the National Key Basic Research Program of China (2015CB921600), the National Natural Science Foundation of China (61625402, 61574076, 11774399), the Collaborative Innovation Center of Advanced Microstructures, Natural Science Foundation of Jiangsu Province (BK20180330 and BK20150055), Fundamental Research Funds for the Central Universities (020414380122, 020414380084), Beijing Natural Science Foundation (Z180008) and the program A/B for Outstanding PhD candidate of Nanjing University(201801A002). K.W. and T.T. acknowledge support from the Elemental Strategy Initiative conducted by the MEXT, Japan, A3 Foresight by JSPS and the CREST (JPMJCR15F3), JST.

# Figures

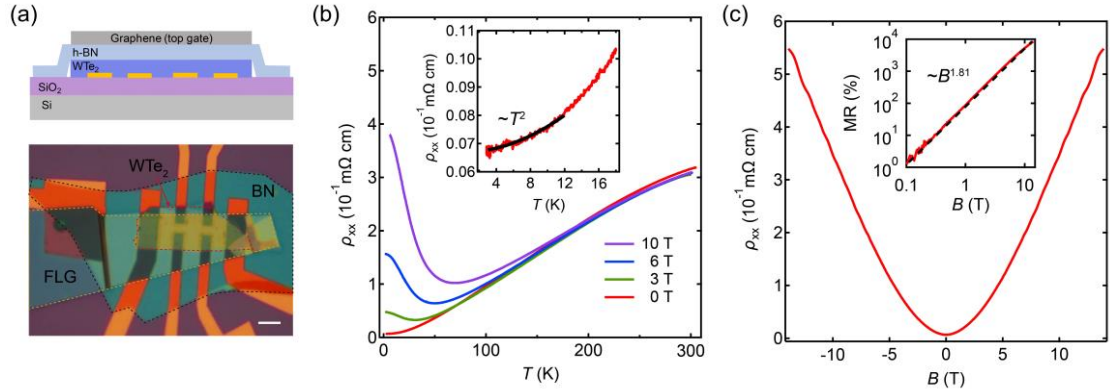

**Figure 1**. Characterization of the thin film device with h-BN as protective layer. (a) Optical image of 10 nm thick WTe$_2$ device, the boundary of the WTe$_2$ film is indicated by the red dashed line. The edges of few-layer graphene and h-BN are indicated by the yellow and black dashed lines, respectively. The length of the white scale bar is 5 μm. The upper panel is the cross-sectional schematic structure of the device. (b) Temperature dependence of resistivity at $V_{tg}$ = -6 V for different magnetic fields. The zero-field $\rho(T)$ curve at low temperature range is fitted by $\rho_{xx}(T) = \rho_0 + aT^2$ (solid black line in the inset). (c) Resistivity versus magnetic field $\rho(B)$ at 1.6 K ($V_{tg}$ = 0 V), the inset shows the fitted result (dashed black line) of the magnetoresistance, which follows MR $\sim B^{1.81}$.

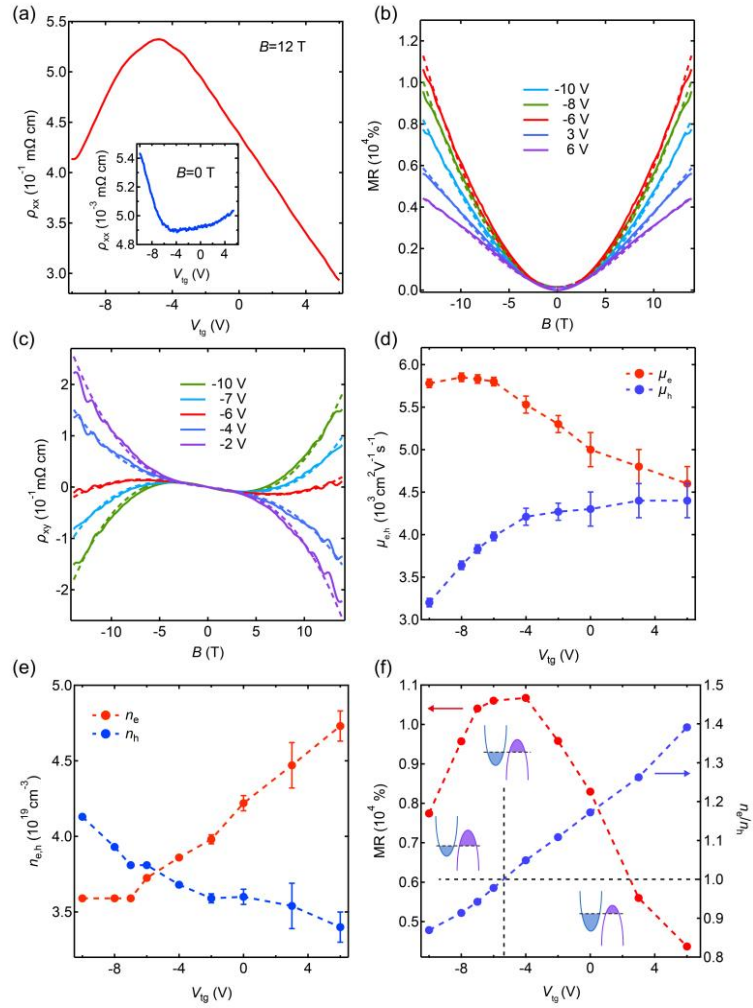

**Figure 2**. Gate-tunable magneto-transport of thin film device at 1.6 K. (a) Gate tunable resistivity with (main panel) and without (inset) magnetic field. (b) and (c) show the gate tunable MR and Hall resistivity curves, respectively. The two-band theory (dashed lines) has been used to fit the MR and Hall resistivity. (d) and (e) are the gate voltage dependence of the extracted mobility and carrier density respectively. The dominant carrier type is distinct at two sides of the crossover point on (e). (f) Comparison of the gate voltage dependent MR and carrier densities ratio. The MR curve reaches the maximum with tuning the carrier concentration through the charge compensation point ($n_e/n_h = 1$). The three schematic bands represent different scenarios for charge distribution in electron (blue) and hole (purple) pockets at different gate voltage ranges.

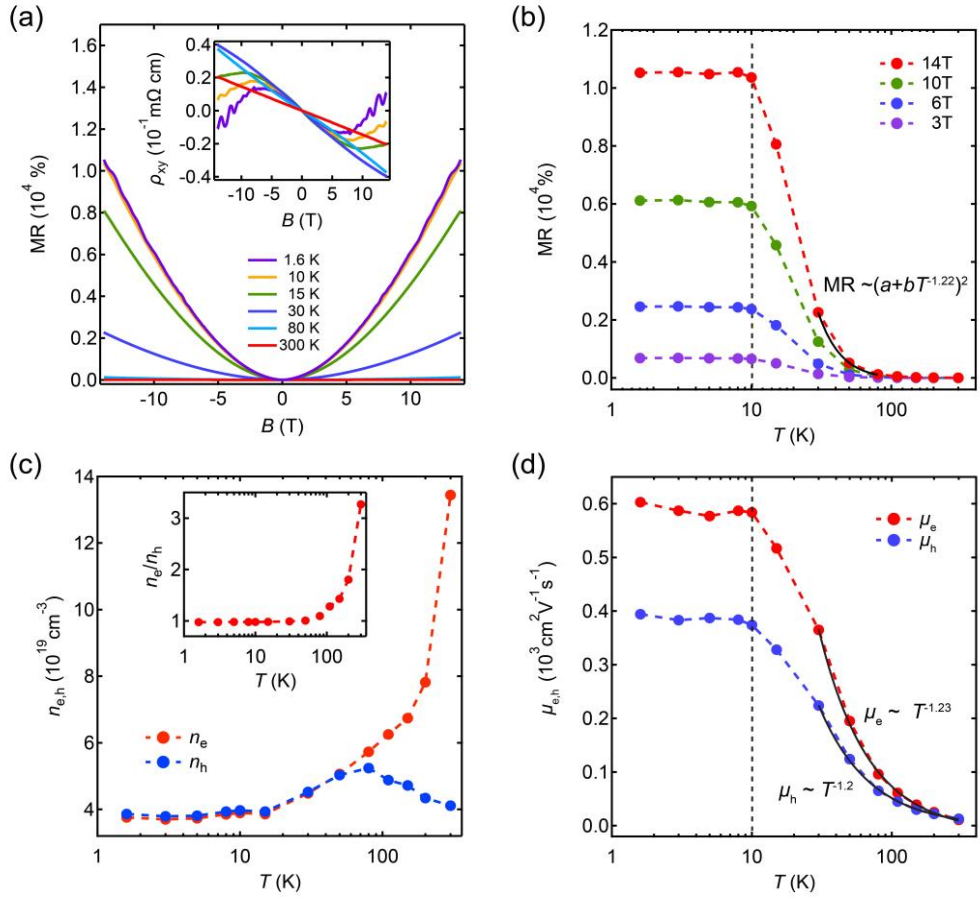

**Figure 3**. Analysis of temperature dependent magneto-transport at $V_{tg}$ = -6 V. (a) MR(*B*) versus magnetic field at different temperatures, the inset shows the corresponding Hall resistivity curves. (b) Temperature dependence of the MR at different magnetic fields. The decrease of MR(*T*) follows ~ $(a + bT^{-1.22})^2$ (the solid black line). (c) Extracted carrier (electron and hole) densities as a function of temperature, the inset is the ratio of corresponding electron and hole densities. (d) Temperature dependence of mobilities for electron and hole. Power law (solid black lines) has been used to fit the experimental data (symbols).